\documentstyle[11pt,newpasp,twoside]{article}
\def\edcomment#1{\iffalse\marginpar{\raggedright\sl#1\/}\else\relax\fi}
\reversemarginpar

\begin{document}
\title{A Search for Disk Emission in Young Brown Dwarfs: $L^{\prime}$-band Observations
of $\sigma$ Orionis and TW Hydrae}
 \author{Ray Jayawardhana}
\affil{Department of Astronomy, University of Michigan, Ann Arbor, MI 48109, U.S.A.}
\author{David R. Ardila}
\affil{Department of Astronomy, University of California, Berkeley, CA 94720, U.S.A.}
\author{Beate Stelzer}
\affil{Max-Planck-Institut f\"ur extraterrestrische Physik, 85741 Garching, Germany}

\begin{abstract}
Studies of disks 
around young brown dwarfs are of paramount importance to our understanding 
of the origin, diversity and early evolution of sub-stellar objects. Here 
we present first results from a systematic search for disk emission in a 
spectroscopically confirmed sample of young objects near or below the 
sub-stellar boundary in a variety of star-forming regions. Our VLT and Keck
$L^{\prime}$-band observations of the $\sigma$ Orionis and TW Hydrae 
associations suggest 
that if a majority of brown dwarfs are born with disks, at least the inner 
regions of those disks evolve rapidly, possibly clearing out within a few 
million years.  
\end{abstract}

\section{Introduction}
The current paradigm of low-mass star formation holds that a young star 
accretes material from a circumstellar disk during the first (few) million  
years of its lifetime. The disk also provides the building material for
planets. Over the past two decades, substantial observational evidence 
has accumulated to support this picture (Mannings, Boss \& Russell 2000).  
However, much of that evidence rests on studies of stars within a relatively 
narrow mass range. In particular, there are few observational constraints
on the formation of objects near or below the sub-stellar boundary. 

If a large fraction of young brown dwarfs indeed harbor disks, the 
implication is that extremely low-mass objects may form via a mechanism 
similar to higher mass stars. Thus, reliable determination of the disk 
frequency as a function of age in young sub-stellar populations is critical 
to our understanding of their origin, diversity and early evolution. 

We have commenced a program to obtain $L^{\prime}$-band data on a large, 
{\it spectroscopically confirmed} sample of young objects near or below the 
sub-stellar boundary in a number of nearby star-forming regions. 
$L^{\prime}$-band 
photometry is much better at detecting disk excess above the photospheric 
emission, and is less susceptible to the effects of disk geometry, than 
measurements at shorter wavelengths.

\section{Observations}
$L^{\prime}$-band photometric observations of six $\sigma$ Ori brown dwarfs 
(B\'ejar et 
al. 2001) were obtained at the ESO Very Large Telescope using ISAAC in 
January 2002 in service mode. We also observed two brown dwarf member
candidates in the TW Hydrae Association, recently found by Gizis (2002), 
in $JHK_s L^{\prime}$ at Keck using NIRC in April 2002. The 2MASS database
contains $JHK_s$ for all objects.

\section{Results and Discussion}
Of the six $\sigma$ Ori sources, only one (SOri 12) shows significant 
$K$-$L^{\prime}$ excess, compared to field objects of the same spectral 
type (from Leggett et al. 2002). Neither of the two TW Hydrae targets 
harbors measurable excess consistent with an optically thick inner disk. 

A large fraction --$\sim$65\%-- of brown dwarf candidates in the Trapezium 
cluster show $K$-band excess (Muench et al. 2001; also see Liu et al., this 
volume). Our results, albeit for a small sample of objects so far, in the 
somewhat older $\sigma$ Ori and TW Hydrae 
associations suggest lower disk fractions. It may be that small grains
in the inner disks have cleared out already by the age of these associations.
If so, brown dwarf disks could deplete rapidly, at timescales comparable
to or smaller than those for T Tauri disks (Jayawardhana et al. 1999;
Haisch et al. 2001). 
$L^{\prime}$-band photometry of larger brown dwarf samples in several young 
clusters, spanning a range of ages, could provide a more definitive answer. 
Further constraints on disk properties await mid- and far-infrared 
observations with SIRTF and SOFIA. 

Gizis (2002) reported strong H$\alpha$ emission (equivalent width $\approx$ 
300 \AA) from one of the TW Hydrae objects, the M8 dwarf 2MASSW 
J1207334-393254, and suggested it could be due to either accretion, flaring, or
chromospheric/coronal activity. Given the lack of substantial 
$L^{\prime}$-band excess in this object, accretion now appears 
less likely as the cause of its strong H$\alpha$ emission. 


\end{document}